\documentclass[lettersize,journal]{IEEEtran}
\usepackage{amsmath,amsfonts}
\usepackage{algorithmic}
\usepackage{algorithm}
\usepackage{array}
\usepackage{subfigure}
\usepackage{textcomp}
\usepackage[version=4]{mhchem}
\usepackage{stfloats}
\usepackage{url}
\usepackage{verbatim}
\usepackage{graphicx}
\usepackage{cite}
\usepackage{amsmath,amsfonts,amssymb}
\usepackage{algorithmic}
\usepackage{algorithm}
\usepackage{array}
\usepackage{textcomp}
\usepackage{stfloats}
\usepackage{lipsum}  
\usepackage{url}
\usepackage{verbatim}
\usepackage{color}
\usepackage{graphicx}
\usepackage{cite}
\usepackage{enumitem}

\DeclareMathOperator\erfc{erfc}

\newcommand{\p}{\mathrm{p}}
\newcommand{\PP}{\mathrm{P}}
\newcommand{\E}{\mathrm{E}}

\newcommand{\Var}{\mathrm{Var}}

\newcommand{\Cov}{\mathrm{Cov}}

\newcommand{\Std}{\mathrm{Std}}
\newcommand{\NR}{N_\mathrm{R}}
\makeatletter
\newcommand{\vastt}{\bBigg@{3}}
\newcommand{\vast}{\bBigg@{4}}
\newcommand{\Vast}{\bBigg@{5}}

\makeatother

\begin{document}

\title{Affinity-Division Multiplexing for \\Molecular Communications with \\Promiscuous Ligand Receptors}

\author{
        Ahmet R. Emirdagi*,
        M. Serkan Kopuzlu*,
        M. Okan Araz*,
        and Murat Kuscu,
       \thanks{*These authors contributed equally.}
       \thanks{The authors are with the Nano/Bio/Physical Information and Communications Laboratory (CALICO Lab), Department of Electrical and Electronics Engineering, Koç University, Istanbul, Turkey (e-mail: \{aemirdagi18, mkopuzlu18, maraz18, mkuscu\}@ku.edu.tr).}
	   \thanks{This work was supported in part by the EU Horizon 2020 MSCA-IF under Grant \#101028935, and by The Scientific and Technological Research Council of Turkey (TUBITAK) under Grant \#120E301.}
}

\maketitle

\begin{abstract}
A key challenge in Molecular Communications (MC) is low data transmission rates, which can be addressed by channel multiplexing techniques. One way to achieve channel multiplexing in MC is to leverage the diversity of different molecule types with respect to their receptor binding characteristics, such as affinity and kinetic binding/unbinding rates. In this study, we propose a practical multiplexing scheme for MC, which is based on the diversity of ligand-receptor binding affinities. This method requires only a single type of promiscuous receptor on the receiver side, capable of interacting with multiple ligand types. We analytically derive the mean Bit Error Probability (BEP) over all multiplexed MC channels as a function of similarity among ligands in terms of their receptor affinities, the number of receptors, the number of multiplexed channels, and the ratio of concentrations encoding bit-1 and bit-0. We investigate the impact of each design parameter on the performance of multiplexed MC system.
\end{abstract}


\section{Introduction}
\IEEEPARstart{I}{nternet} of Bio-Nano Things (IoBNT) describes the smart and cooperative networks of heterogeneous `bio-nano things', such as natural and artificial biological cells, micro/nano robots, and functional nanoparticles, and is promising for expanding the capabilities of emerging nano and biotechnologies (e.g., synthetic biology) by integrating them with information and communication technologies (ICT) \cite{kuscu2021internet}. Harnessing the power of the \emph{collective}, envisioned applications of IoBNT include therapeutic and diagnostic applications (e.g., smart drug delivery, intrabody continuous-health monitoring), collective sensing, biocomputing, and biomanufacturing\cite{akyildiz2015internet,akan2016fundamentals}.  Bio-inspired Molecular Communications (MC) has emerged as the most promising solution to address the communication problem in complex biochemical application environments of IoBNT. Past decade has witnessed a surge in theoretical studies to understand the performance limits of this unconventional communication modality and to develop practical MC techniques, such as modulation, detection, and channel estimation, that can meet the performance requirements of IoBNT applications. 

As revealed by many theoretical and experimental studies, a key problem in MC is the extremely low data transmission rates, which is partly due to the slow propagation of molecules in fluidic channels and slow reaction kinetics at the channel-receiver interface \cite{kuscu2021fabrication}. Channel multiplexing can improve the data transmission capacity by allowing multiple messages to be transmitted simultaneously over the same physical channel. Multiplexing can also enable multi-access in MC networks which are crowded by several transmitter and receiver pairs. Several strategies have been proposed for channel multiplexing in MC. In \cite{freq_div}, the authors proposed Frequency Division Multiplexing (FDM) scheme for MC exploiting the longitudinal carrier wave properties of propagating MC signals. Time Division Multiple Access (TDMA) has been studied for MC in \cite{TDMA, rudsari2019drug, rudsari2021tdma}. Although \cite{rudsari2021tdma} considers a receiver with ligand receptors, none of the multiplexing schemes proposed in the literature exploit the statistics of ligand-receptor binding interactions to increase the utilization of the MC channel. 

Ligand-receptor interactions are fundamental to sensing and communication in biological cells by serving as a selective and sensitive interface with the complex and dynamic biochemical environment. However, ligand receptors are not perfect in terms of selectivity, and may bind to various types of ligands, albeit with differing affinities. This phenomenon, known as \emph{receptor promiscuity}, resulting in cross-talk of ligands is shown to enable unique opportunities for developing practical MC techniques that can be implemented in bio-synthetic MC transceivers \cite{kuscu2021detection, kopuzlu2022capacity}. For example, in \cite{kuscu2019channel}, we showed that the bound and unbound time durations of a single promiscuous receptor can inform about the individual concentrations of multiple types of ligands co-existing in the channel. In the present study, we follow a similar strategy to design an MC channel multiplexing scheme, i.e., \emph{affinity-division multiplexing (ADM)}, where the transmitter encodes the messages into the concentrations of different types of ligands based on Binary-Concentration Shift Keying (B-CSK) modulation, which are then released into the channel simultaneously. 

In the investigated scenario, the receiver estimates the concentrations of transmitted individual ligand types and performs the detection using only a single type of receptors based on their bound and unbound time duration statistics. In our previous work \cite{kuscu2021detection, kuscu2019channel}, we explored the feasibility of implementing such receiver architecture in bio-synthetic devices (i.e., engineered or artificial cells). We demonstrated that the ligand concentration estimation scheme, which is a key component of the proposed multiplexing method, can be realized by custom-designed chemical reaction networks (CRNs) and synthetic receptors. These synthetic receptors employ a modified version of the biological kinetic proofreading (KPR) mechanism, which is responsible for highly specific antigen recognition of T-cells in the immune system \cite{mckeithan1995kinetic}.

To evaluate the performance of the proposed multiplexing scheme, we analytically derive the mean Bit Error Probability (BEP) across all multiplexed MC channels. The evaluation takes into account the varying number of receptors, number of multiplexed channels, the similarity among the utilized ligand types in terms of their affinity for receptors, and the ratio of concentrations encoding bit-1 and bit-0.

The remainder of this paper is organized as follows. In Section \ref{sec:estimation}, we present the derivation of the ligand concentration estimation scheme. In Section \ref{sec:sys_model}, we provide a detailed description of the considered system model, and derive the analytical expression for mean BEP in the multiplexed MC channels. This expression is then used for the numerical evaluation of the multiplexing performance in Section \ref{sec:numerical}. Finally, we conclude the paper in Section \ref{sec:conclusion}.

\section{Estimation with Ligand Receptors}
\label{sec:estimation}

In case of monovalent ligands, ligand-receptor binding reactions can be modelled as a two-state continuous-time stochastic process with the states corresponding to the bound (B) and unbound (U) states of a receptor. In case of a single ligand type, this reaction can be represented as
\begin{equation}
\ce{\mathrm{U}  <=>[{c_\mathrm{L}(t) k^+}][{k^-}] \mathrm{B}},
\label{equilibrium}
\end{equation}
where, $c_\mathrm{L}(t)$ is the ligand concentration in the vicinity of the receptor, $k^+$ and $k^-$ are the binding and unbinding rates of the ligand-receptor pair, respectively. Under equilibrium conditions, where ligand concentration can be assumed stationary, and in the case that $M$ different types of ligands co-exist in the channel with concentrations $\boldsymbol{c} = [c_1, ..., c_i, ..., c_M]$, the probability of finding a receptor in the bound state can be given as \cite{kuscu2021detection}
\begin{equation}
   \p_\mathrm{B}=\frac{\sum_{i = 1}^{M} c_i/K_{\mathrm{D},i}}{1+ \sum_{i = 1}^{M} c_i/K_{\mathrm{D},i}},
   \label{eq:prob_binding_mixture}
\end{equation}
where $K_{\mathrm{D},i} = k^-_i/k^+_i$ is the dissociation constant of the receptor for the $i^\text{th}$ ligand type. From this relation, the individual ligand concentrations $c_i$ cannot be determined unequivocally. However, as shown in \cite{kuscu2021detection}, the individual ligand concentrations can be estimated accurately using the bound and unbound time duration statistics of receptors, which are informative of the concentration ratio of different ligand types, and the total ligand concentration, respectively. Given that the receiver has $N_\mathrm{R}$ receptors which are independent from each other and that the binding and unbinding processes are ergodic, the log-likelihood of observing a set of receptor bound time durations $\{\tau_\mathrm{b}\}_{N_\mathrm{R}}$, and a total unbound time duration over all receptors $T_\mathrm{u} = \sum_{i=1}^{N_\mathrm{R}} \tau_{\mathrm{u},i}$ can be given as follows
\begin{align}
\nonumber
    \mathcal{L}(\{\tau_\mathrm{b}\}_{N_\mathrm{R}}|\boldsymbol{\alpha)} &= \sum_{i=1}^{\NR} \ln \p(\tau_{\mathrm{b},i}),\\
    \mathcal{L}(T_\mathrm{u}|c_\mathrm{tot}) &= \NR\ln(c_\mathrm{tot}) - k^+c_\mathrm{tot}T_\mathrm{u},
\end{align}
where $c_\mathrm{tot} = \sum_{i=1}^M c_i$ is the total ligand concentration in the vicinity of the receptors, $\boldsymbol{\alpha} = [\alpha_1, ..., \alpha_i = c_i/c_\mathrm{tot}, ..., \alpha_M]$ is the vector of ligand concentration ratios, $\p(\tau_{\mathrm{b},i})$ is the probability of observing a bound time duration in the $i^\text{th}$ receptor, which is a mixture of exponential distributions defined by the unbinding rates of co-existing ligand types, $k^-_i$ \cite{kuscu2019channel}.

By solving ${\partial \mathcal{L}(T_\mathrm{u}|c_\mathrm{tot})}/{\partial c_\mathrm{tot}}=0$, we can derive the Maximum Likelihood (ML) estimator for $c_\mathrm{tot}$ as $\hat{c}_\mathrm{tot}=\frac{\NR}{k^{+} T_\mathrm{u}}$, where $T_\mathrm{u}$ is gamma distributed being the sum of $\NR$ independent exponential random variables (i.e., $\tau_{\mathrm{u},i}$). This makes its reciprocal, $1/T_\mathrm{u}$, an inverse gamma-distributed random variable with the mean $\E[1/T_\mathrm{u}]=\frac{k^{+} c_\mathrm{tot}}{\NR-1}$. Hence, the mean of the total ligand concentration estimator becomes $\E[\hat{c}_\mathrm{tot}]= (\NR/k^{+}) \times \E[1/T_\mathrm{u}]= c_\mathrm{tot} \times (\NR/(\NR-1))$, which indicates that the estimator is biased unless $\NR$ is very large. Instead, we prefer using an unbiased version of estimator, which is obtained with simple modification as follows
\begin{equation}
\label{eq:success_prob_csk}
   \hat{c}_\mathrm{tot}=\frac{\NR-1}{k^{+} T_\mathrm{u}},
\end{equation}
with the mean and variance
\begin{align}
\nonumber
    \E[\hat{c}_\mathrm{tot}] &=c_\mathrm{tot}, \\
    \Var[\hat{c}_\mathrm{tot}]&=\frac{c_\mathrm{tot}^2}{\NR-2} 
\hspace{2mm} \text{for} 
\hspace{2mm} \NR>2.
\end{align}

To estimate the concentration ratios of individual ligand types, $\alpha_i$, we use a method that involves separating receptor bound times into groups based on intervals determined by the unbinding rate of the ligands, rather than sampling the exact bound time duration \cite{kuscu2019channel}. The method estimates ligand concentration ratios by counting the number of binding events within specific time intervals. The number of intervals is equal to the number of distinct ligand types, $M$, and they are defined by time thresholds proportional to the inverse of the unbinding rates of the different ligand types, i.e.,
\begin{equation}
    T_i = v/k^-_i
    \hspace{2mm} \text{for}  \hspace{2mm} i \in \{1,\dots, M-1\}.
\end{equation}
Here, the ligand types are indexed in the order of increasing affinity for receptors, i.e., decreasing $k^-_i$. The proportionality constant, $v$, can be optimized for better estimation performance. In this study, we set $v = 3$ in accordance with \cite{kuscu2019channel}. The probability of observing a binding event that falls in a region between two time thresholds is then given as
\begin{equation}
\nonumber
    \p_{l} = \int_{T_{l-1}}^{T_{l}} \p({{\tau_\mathrm{b}}'}) \,d{{\tau_\mathrm{b}}'} = \sum_{i=1}^M\alpha_i \left(  e^{-{(k_j^-}{T_{i-1}})} - e^{-{(k_j^-}{T_i})}\right)  ,
\end{equation}
with $T_\mathrm{0} = 0$ and $T_{M} = +\infty$. Equivalently, we can write
\begin{equation}
    \boldsymbol{\p} = \boldsymbol{S}\boldsymbol{\alpha},
\end{equation}
where $\boldsymbol{\p}$ is $(M \times 1)$  probability vector, and $S$ is a $(M \times M)$ matrix with elements
\begin{equation}
    s_{i,j} = e^{-{(k_j^-}{T_{i-1}})} - e^{-{(k_j^-}{T_i})}.
\end{equation}
The number of independent binding events falling in a time interval follow a binomial distribution with mean and variance
\begin{align}
    \boldsymbol{\E}[\boldsymbol{n}] &= \boldsymbol{\p}\NR, \\
    \boldsymbol{\Var}[\boldsymbol{n}] &= (\boldsymbol{\p} \odot (\boldsymbol{1-\p})) \NR,
\end{align}
where $\boldsymbol{n}$ is $(M \times 1)$ vector of number of binding events that fall in each interval, and $\odot$ is the Hadamard product. We use Method of Moments to obtain an estimator for concentration ratios from the number of binding events as follows 
\begin{equation}
    \boldsymbol{n} = \boldsymbol{\hat{\p}}\NR = \boldsymbol{S} \boldsymbol{\hat{\alpha}} \NR,
\end{equation}
where hat represents the estimated version of the parameters. By leaving  $\boldsymbol{\hat{\alpha}}$ alone and using $\boldsymbol{W} = \boldsymbol{S^{-1}}$, we obtain the ratio estimator
\begin{align}
    \boldsymbol{\hat{\alpha}} = \left(\frac{1}{\NR}\right) \boldsymbol{Wn},
\end{align}
with the mean and variance
\begin{align}
\nonumber
\label{eq:mean_var_subop}
    \boldsymbol{\E}[\boldsymbol{\hat{\alpha}}] &= \frac{1}{\NR} \boldsymbol{W} \bold{E}[\boldsymbol{n}] = \boldsymbol{W}\bold{p} = \boldsymbol{S^{-1}}\bold{p} = \boldsymbol{\alpha}, \\
    \Var[\hat{\alpha}_{l}] &= \frac{1}{\NR^2} \sum_{i=1}^{M} \sum_{j=1}^{M} w_{l,i} w_{l,j} \Cov[n_i,n_j],
\end{align}
where $l \in \{1, \dots, M\}$, and the covariance function is 
\begin{equation}
    \mathrm{Cov}[n_i,n_j] = \left\{
	\begin{array}{ll}
		\Var[n_i],  & \mbox{if } i = j, \\
		-\p_i \p_j \NR, & \mbox{otherwise.} 
	\end{array}
\right.
\end{equation}

By combining the ratio estimator with the total ligand concentration estimator, we can obtain an ML estimate for the concentration of individual ligand types as follows
\begin{equation}
    \hat{\boldsymbol{c}} = \hat{c}_\mathrm{tot} \boldsymbol{\hat{\alpha}},
\end{equation}
with the mean and variance given by
\begin{align}
\nonumber
    \boldsymbol{\E}[\hat{\boldsymbol{c}}] &= \E[\hat{c}_\mathrm{tot}]\boldsymbol{\E}[\boldsymbol{\hat{\alpha}}] = c_\mathrm{tot}\boldsymbol{\alpha}=\boldsymbol{c}, \\
    \nonumber
    \boldsymbol{\Var}[\hat{\boldsymbol{c}}] &= \Var[\hat{c}_\mathrm{tot}]\boldsymbol{\Var}[\boldsymbol{\hat{\alpha}}] +\Var[\hat{c}_\mathrm{tot}] (\boldsymbol{\E}[\boldsymbol{\hat{\alpha}}] \odot \boldsymbol{\E}[\boldsymbol{\hat{\alpha}}]) \\ 
    &\hspace{10mm} +\boldsymbol{\Var}[\boldsymbol{\hat{\alpha}}]\E[\hat{c}_\mathrm{tot}]^2,
\end{align}
indicating that the estimator $\boldsymbol{\hat{c}}$ is also unbiased. The distribution of individual ligand concentration estimates can be approximated as Gaussian distributed, i.e., $\hat{c}_i \sim \mathcal{N}\left(\E[\hat{c}_i],\Var[\hat{c}_i]\right)$, for sufficiently high $\NR$, corresponding to the number of independent samples taken from the independent receptors \cite{kuscu2021detection}. 

\section{Affinity-Division Multiplexing}
\label{sec:sys_model}
We study an MC system with a single pair of bio-synthetic MC transmitter and receiver. Information is encoded into the concentration of a different type of ligand for each channel using B-CSK. Receiver estimates the individual ligand concentrations for each channel utilizing the bound and unbound time statistics of receptors to decode the transmitted symbol for each channel. The following settings and assumptions describe the considered MC system:
\begin{itemize}[leftmargin=*]
     \item The number of multiplexed channels is equal to the number of different ligand types utilized by the transmitter, i.e., $N_\mathrm{C} = M$.
     
    \item The transmitter releases all types of ligands to the channel at the same time instant as an impulse, i.e., $\boldsymbol{x}(t) = \boldsymbol{N}_\mathrm{tx}\delta(t)$, where $\boldsymbol{N}_\mathrm{tx}$ is $(M \times 1)$ vector of the form $\boldsymbol{N}_\mathrm{tx} = \boldsymbol{s} N_1 + \boldsymbol{\tilde{s}} N_0$, with $\boldsymbol{s}$ refers to the $(M \times 1)$ vector containing the transmitted bits for each channel, i.e., $s_i \in \{0,1\}$, and the $\boldsymbol{\tilde{s}}$ denotes the negation of that vector. In the above notation, $N_0$ and $N_1$ denotes the released number of molecules to encode bit-0 and bit-1, respectively. Both $N_0$ and $N_1$ are the same across all channels.
    
    \item The transmitter has equal probability of sending bit-0 or bit-1 in each multiplexed channel.
    
    \item The receiver has only a single type of receptors that are uniformly distributed on its surface. These receptors are identical, non-interacting, and independent, and are exposed to the same total ligand concentration. We disregard any types of regulation of receptors by bio-synthetic cells, such as clustering, which could result in receptor correlations \cite{roob2016cooperative}. Furthermore, we neglect correlations between neighboring receptors that may arise from local ligand concentration gradients due to diffusion or ligand binding to receptors, assuming that ligands are abundant in the vicinity of receptors and that receptors are sufficiently distant from each other. This assumption holds true as long as $N_\mathrm{R} r_\mathrm{R} > r_\mathrm{RX}$, with $N_\mathrm{R}$ and $r_\mathrm{R}$ representing the number and radius of receptors, respectively, and $r_\mathrm{RX}$ denoting the radius of the receiver \cite{bialek2005physical}. Considering that typical cell membrane receptors (e.g., GPCRs) have estimated diameters ranging between $4-10$nm, a biosynthetic MC receiver with a diameter of $10\mu$m would satisfy this constraint.
    
    \item All ligand types have the same diffusion coefficient and bind to the receptors with an identical binding rate. This assumption is considered realistic when ligand-receptor interactions are diffusion-limited \cite{kuscu2019channel}.

    \item Ligand concentrations near the receptors do not change within the sampling window, due to the low-pass characteristics of the MC channel, and any variations in ligand concentrations caused by the binding reactions are considered to be insignificant when measuring the bound time intervals. 

    \item The receiver knows the unbinding rates of the ligands utilized for multiplexing. The degree of similarity between two successive ligand types is quantified by the similarity parameter $\gamma_i$, which is defined as the ratio of their unbinding rates, i.e., $\gamma_i = k_i^-/k_{i+1}^-$. We assume that $\gamma_i$ is equal for all $i \in \{1, \dots, M-1\}$.
 
\end{itemize}
    
\underline{\textbf{Received Signal:}} The actual received signal can be modeled using the channel impulse response (CIR) of the MC channel. For both transmitter and receiver being stationary, and assuming that the receiver takes its samples at the peak time of the received concentration after the signaling, i.e., $\tau_\mathrm{peak}$, CIR can be given as
\begin{equation}
    h(\tau_\mathrm{peak}) = \left(\frac{{2 \pi  r^2}}{3}\right)^{-3/2} \mathrm{exp} \left(-{\frac{3}{2}} \right),
\end{equation}
where $r$ is the transmitter-receiver distance \cite{ahmadzadeh2018stochastic}. Accordingly, the actual received signal in the vicinity of the receptors can be represented as
\begin{equation}
    \boldsymbol{c} = h(\tau_\mathrm{peak}) \boldsymbol{N}_\mathrm{tx},
\end{equation}
with the elements, $c_i$, representing now the actual received concentration in the $i^\text{th}$ channel, where the $i^\text{th}$ ligand type is utilized. 

The estimated received signal, which we refer to simply as the received signal from here on, is the array of estimates of actual received ligand concentrations in the individual channels, i.e., $\boldsymbol{\hat{c}} = [\hat{c}_1, ..., \hat{c}_i, ..., \hat{c}_M]$. The ligand concentration estimate in the $i^\text{th}$ channel, however, is influenced by the transmitted bits not only in the $i^\text{th}$ channel but in all multiplexed channels, $\boldsymbol{s}$, due to co-dependent variations in $\boldsymbol{\alpha}$ and $c_\mathrm{tot}$. Given that there are $2^M$ distinct possibilities for bit transmissions across all channels, when conditioned on a particular bit $s_i$ in the $i^\text{th}$ channel, the mean and the variance of the ligand concentration estimate in the $i^\text{th}$ channel can be obtained by applying the laws of total expectation and variance, i.e., 
\begin{align}
\E[\hat{c}_i | s_i] &= \frac{1}{2^{M-1}} \sum_{\boldsymbol{s^i} \in \boldsymbol{V_{M-1}}} \E[\hat{c}_i | s_i, \boldsymbol{s^i}], \\ \nonumber
\Var[\hat{c}_i | s_i]&= \frac{1}{2^{M-1}} \sum_{\boldsymbol{s^i} \in \boldsymbol{V_{M-1}}} \Var[\hat{c}_i | s_i, \boldsymbol{s^i}],
\end{align}
where $\boldsymbol{s^i} = [s_1, ..., s_{i-1}, s_{i+1}, ... s_M]$ is the $\boldsymbol{s}$-vector excluding the element $s_i$, and $\boldsymbol{V_{M-1}}$ is the set of all binary $(M-1)$-tuples, each having a probability of $1/2^{M-1}$. 

\underline{\textbf{Bit Error Probability:}}
Assuming that the receiver employs ML detection, the decision rule for the $i^\text{th}$ channel becomes
\begin{equation}
    \hat{s}_i = \underset{s_i \in \{0,1\}}{\operatorname{arg~max}}~\PP(\hat{c}_i | s_i),    
\end{equation}  
which can be simplified by defining a decision threshold: $\hat{c}_i \underset{\hat{s}_i = 0} {\overset{\hat{s}_i = 1}{\gtrless}} \lambda_i$.
For Gaussian distributed statistics of the received signal, i.e., $\PP(\hat{c}_i | s_i) \sim \mathcal{N}\left(\E[\hat{c}_i | s_i], \Var[\hat{c}_i | s_i] \right)$, the optimal decision threshold, $\lambda_i$, which minimizes BEP in the $i^\text{th}$ channel can be computed as
\begin{align} \label{threshold2}
\lambda_i  &= \Gamma^{-1}_i  \Biggl( \Var[\hat{c}_i|1] \E[\hat{c}_i|0] \\ \nonumber
 &~~~ - \Var[\hat{c}_i|0] \E[\hat{c}_i|1]  + \Std[\hat{c}_i|1]\Std[\hat{c}_i|0] \\ \nonumber
&~~~ \times \sqrt{\bigl(\E[\hat{c}_i|1] - \E[\hat{c}_i|0] \bigr)^2+2 \Gamma_i  \ln{\frac{\Std[\hat{c}_i|1]}{\Std[\hat{c}_i|0]}}} ~\Biggr), 
\end{align}
where $\Gamma_i = \Var[\hat{c}_i|1] - \Var[\hat{c}_i|0]$, and $\Std[.] = \sqrt{\Var[.]}$ is the standard deviation. The corresponding BEP in the $i^\text{th}$ channel, $\PP_{\mathrm{e},i}$, can be calculated as follows

\begin{align}
    \PP_{\mathrm{e},i} &= \frac{1}{2} \bigg[ \PP(\hat{s}_i = 1 | s_i = 0) + \PP(\hat{s}_i = 0 | s_i = 1)  \bigg] \\ \nonumber
    &=\frac{1}{4}\Biggl[\erfc \Biggl(\frac{\lambda_{i} - \E[\hat{c}_i|0] }{\sqrt{2 \Var[\hat{c}_i|0] }}\Biggr) +  \erfc \Biggl(\frac {\E[\hat{c}_i|1]  - \lambda_{i}}{\sqrt{2 \Var[\hat{c}_i|1]} }\Biggr)\Biggr],
\end{align} 
where $\mathrm{erfc}(z)= \frac{2}{\sqrt{\pi}} \int_{z}^{\infty} e^{-y^2}dy$ is the complementary error function. Overall error performance of ADM can be represented by the mean BEP across all multiplexed channels, i.e., 
\begin{equation}
    \bar{\PP}_\mathrm{e} = \frac{1}{M}\sum_{i=1}^{M} \PP_{\mathrm{e},i}.
\end{equation}

\section{Numerical Results}
\label{sec:numerical}
We evaluate the error performance of ADM in relation to the number of receptors, number of multiplexed channels, and similarity of ligands, and the ratio of number of transmitted molecules for bit-1 and bit-0. The default values of the design parameters used in the analysis are as follows: $\NR = 1000$, $N_\mathrm{C} = 5$, $\gamma = 5$, $r = 20 \mu m$, and $N_1 = 5 \times N_0 = 10^6$.

\subsubsection{\textbf{Effect of Number of Receptors}}
Number of receptors corresponds to the number of independent samples taken for the estimation of received ligand concentrations. As is seen in Fig. \ref{fig:receptor}, increasing number of samples improves the estimation accuracy, thereby decresing the mean BEP over all multiplexed channels.

\subsubsection{\textbf{Effect of Ligand Similarity}}
\begin{figure*}[t!]
  \centering
  \subfigure[]{\label{fig:receptor}\includegraphics[width=0.4\linewidth]{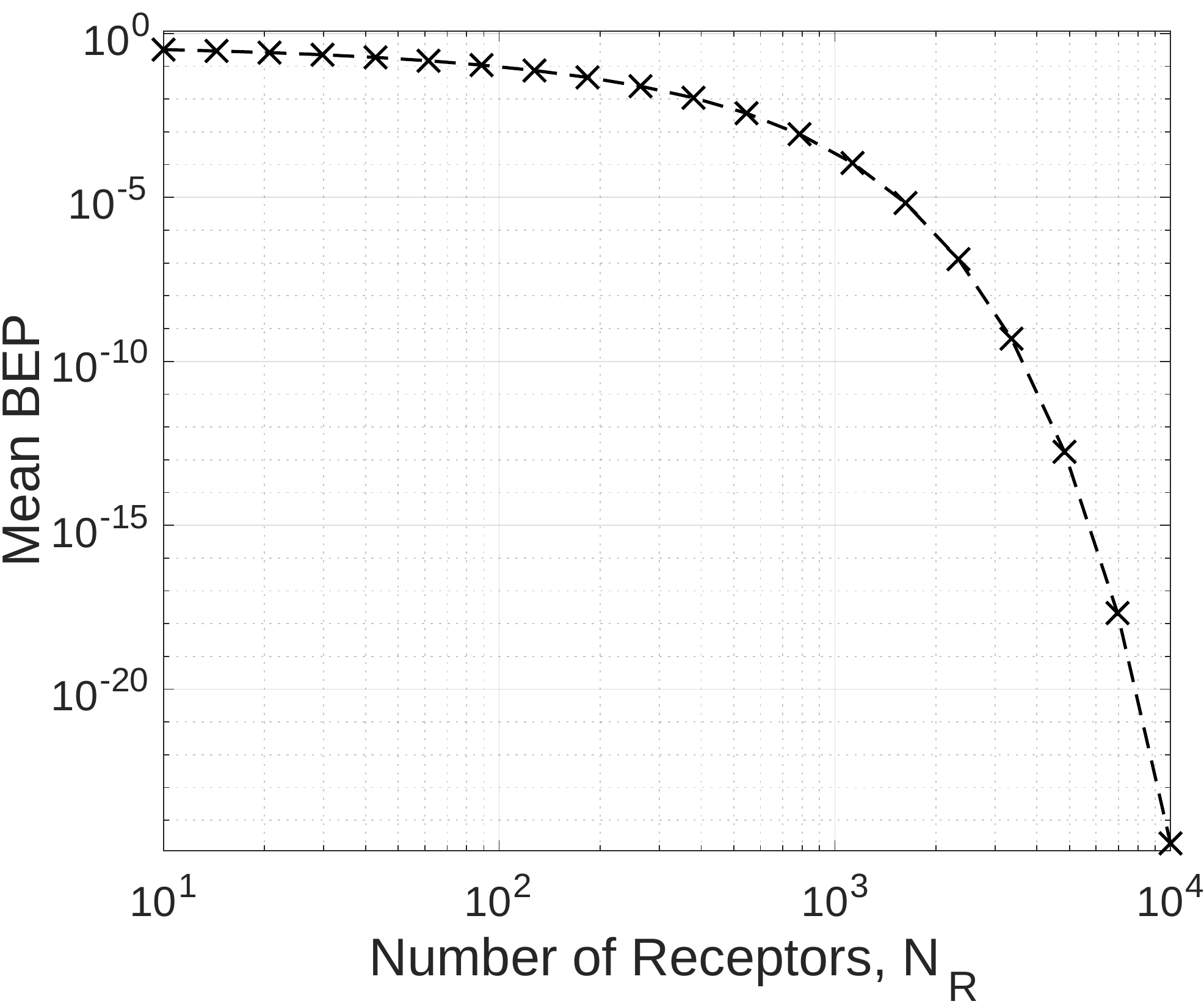}}\quad
    \subfigure[]{\label{fig:similarity}\includegraphics[width=0.4\linewidth]{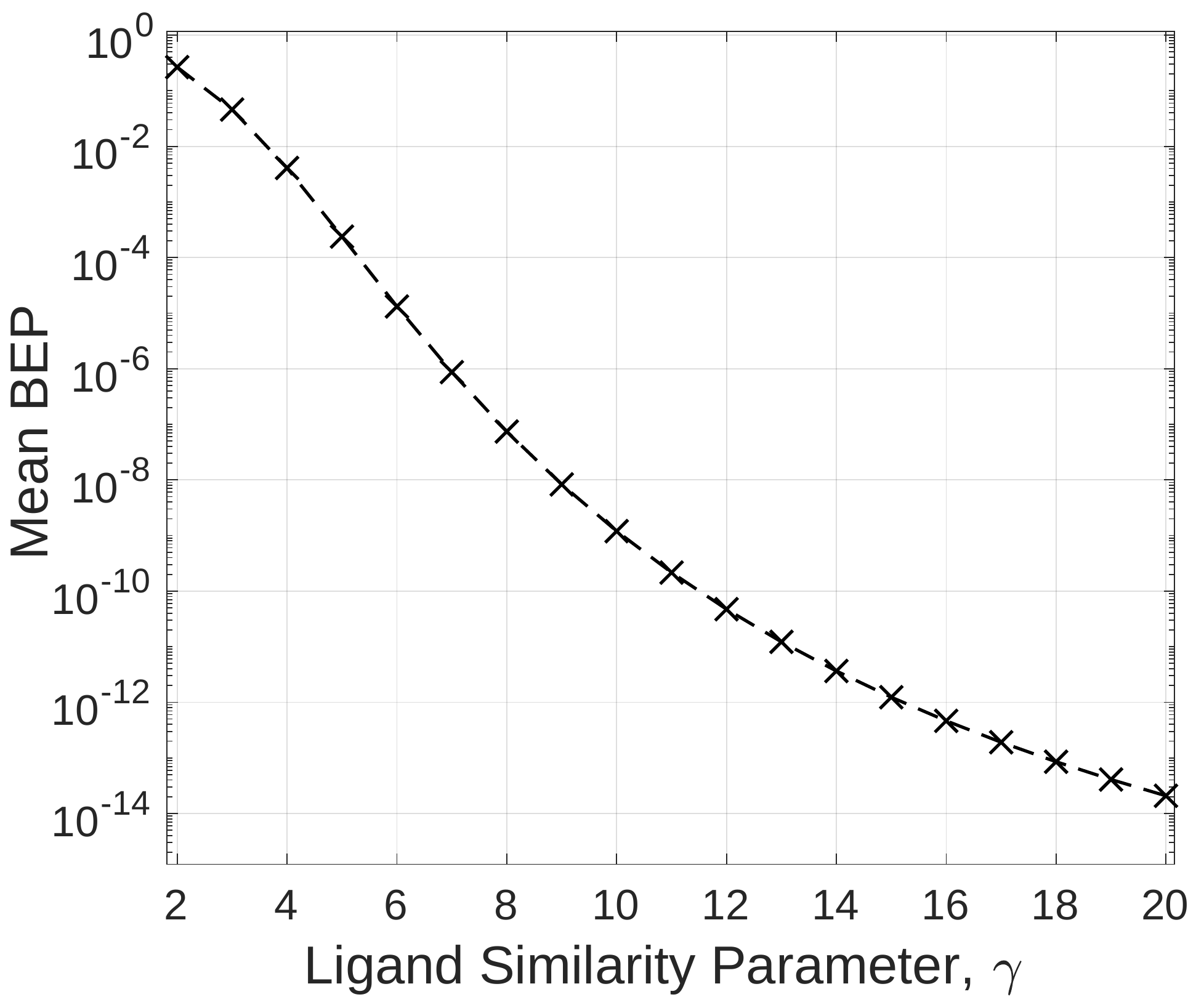}}\quad
   \subfigure[]{\label{fig:channel}\includegraphics[width=0.4\linewidth]{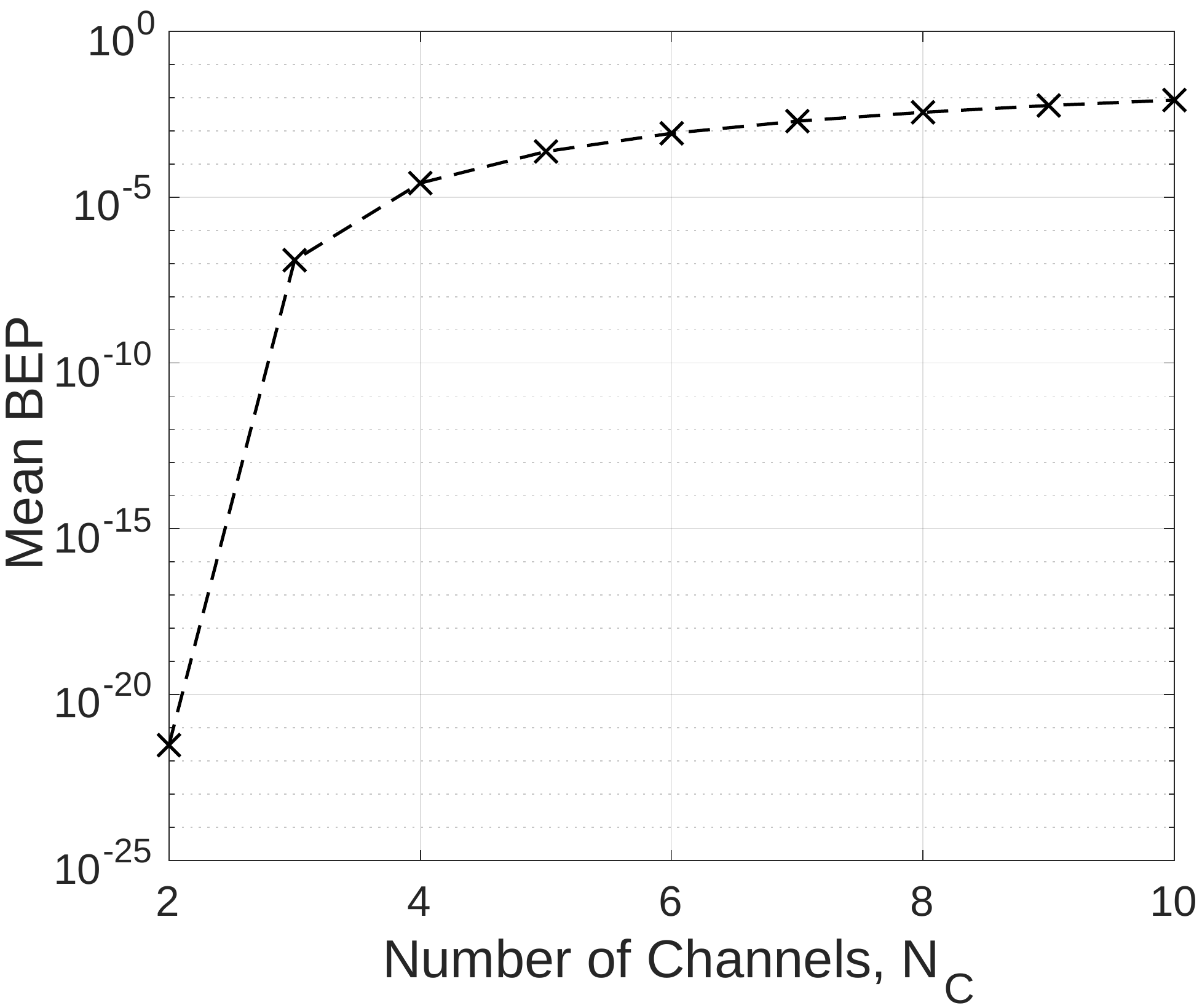}}\quad
   \subfigure[]{\label{fig:bit1bit0}\includegraphics[width=0.4\linewidth]{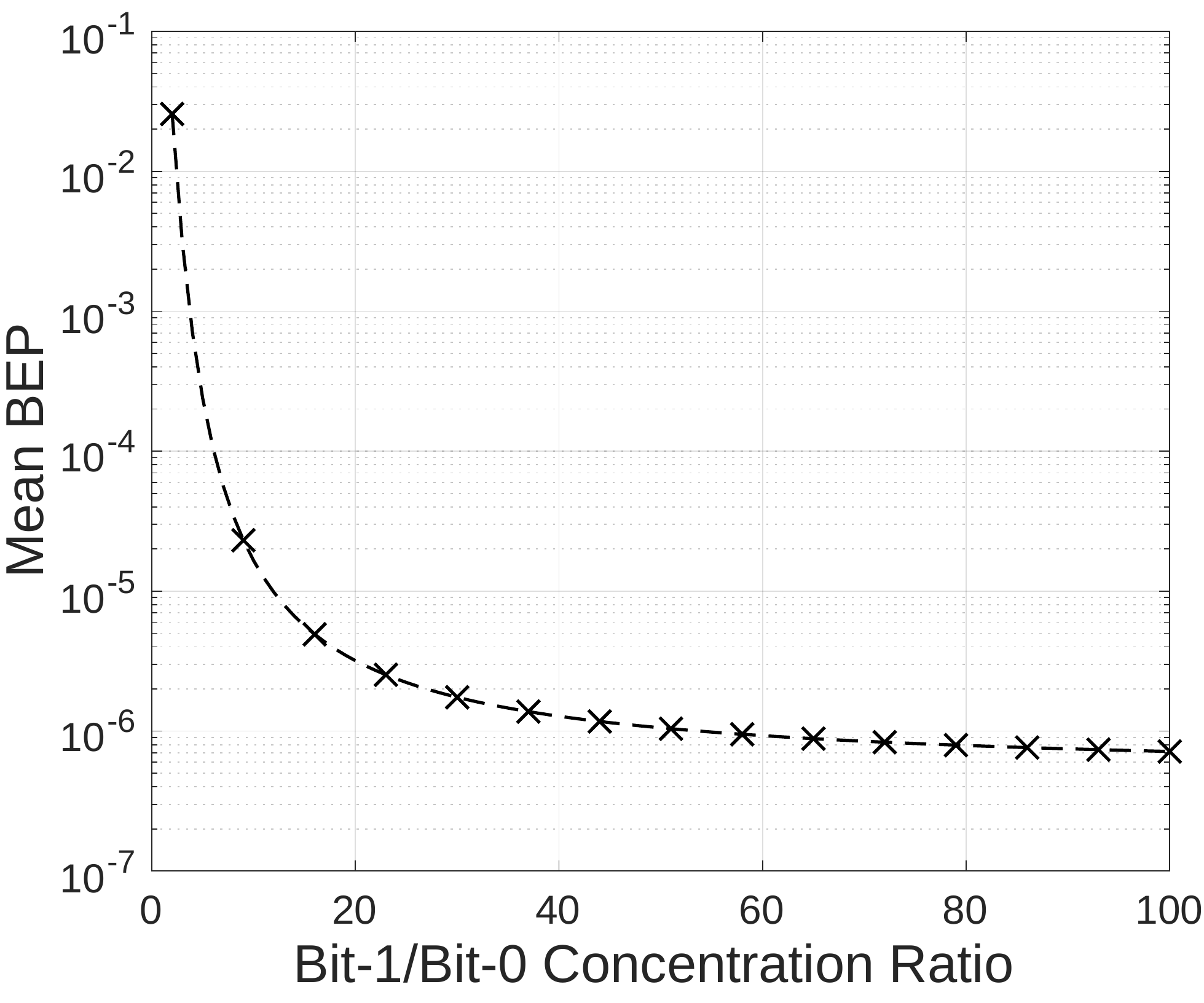}}\quad
   \caption{Error performance of ADM in terms of mean BEP across all multiplexed channels, i.e., $\bar{\PP}_\mathrm{e}$,  for varying (a) number of receptors, (b) degree of similarity among ligands, (c) number of multiplexed channels, and (d) bit-1/bit-0 concentration ratio.}
 \end{figure*}

The similarity among ligands, quantified by the similarity parameter $\gamma$, has a significant impact on the error performance. As the ligands become more similar to each other, the estimator's performance of discriminating between different types of ligands deteriorates. Therefore, we observe a lower BEP for higher $\gamma$, which indicates lower similarity, as shown in Fig. \ref{fig:similarity}. 

\subsubsection{\textbf{Effect of Number of Channels}}
The number of multiplexed channels is equal to the number of different types of ligands utilized for multiplexing. Increasing number of ligand types also increases the variance of the concentration estimator. Therefore, mean BEP of multiplexed channels increases with the number of channels, as is seen in Fig. \ref{fig:channel}. 

\subsubsection{\textbf{Effect of Bit-1/Bit-0 Concentration Ratio}}
The ratio of transmitted ligand concentrations corresponding to bit-1 and bit-0 determines how easily the receiver can distinguish between these two symbols. As the ratio increases, the concentrations encoding different bits become more distinct from each other, improving the receiver's capability to accurately differentiate between them. This leads to a lower mean BEP across all channels as the ratio increases. The results presented in Fig. \ref{fig:bit1bit0} clearly align with this expectation.

\section{Conclusion}
\label{sec:conclusion}

In this paper, we proposed a practical MC multiplexing scheme that leverages the promiscuity of biological receptors, which can bind to different types of ligands with varying affinities. The performance of the proposed affinity-division multiplexing method was evaluated based on several design parameters. Our findings revealed that the degree of similarity among ligand types, in terms of their receptor affinity, significantly impacts the error performance. Overall, the proposed multiplexing scheme demonstrates promising performance for increasing data transmission rates in MC channels, also hinting at its potential in enabling novel multiple access schemes for MC nanonetworks. Future research directions include the validation of our findings using realistic particle-based spatial stochastic simulations. Simulations can facilitate a comprehensive examination of the efficacy of the proposed framework within practical scenarios, while accounting for the effects of diffusion-influenced correlations between receptors for various receiver sizes and geometries. Moreover, exploring non-equilibrium estimation of individual ligand concentrations can be valuable in relaxing the equilibrium assumption, which may not hold true for long receptor sampling windows.

\bibliography{references}
\bibliographystyle{IEEEtran}

\newpage

\vfill

\end{document}